\documentclass[journal=jacsat,manuscript=article, natbib=true]{achemso}
\setcitestyle{numbers,square}
\usepackage[version=3]{mhchem}
\usepackage{adjustbox}
\usepackage{threeparttable}
\usepackage{booktabs}
\usepackage[font=bf]{caption}
\usepackage{graphicx}
\usepackage{subcaption}  
\usepackage{rotating}
\usepackage{bm}
\usepackage{xcolor}
\usepackage[normalem]{ulem}

\SectionNumbersOn

\usepackage[colorlinks=true,
            linkcolor=blue,
            citecolor=magenta,
            urlcolor=blue]{hyperref}

\author{Amrita Manna}
\author{Sudipta Chakraborty}
\affiliation{ 
Department of Chemistry, Indian Institute of Technology Bombay, Mumbai 400076, India
}%
\author{Alexander~B.~Trofimov}
\affiliation{
Laboratory of Quantum Chemical Modeling of Molecular Systems, Irkutsk State University, Karl Marx Str. 1, 664003 Irkutsk, Russia
}%
\alsoaffiliation{
Federal Research Center, A.E. Favorsky Irkutsk Institute of Chemistry, Siberian Branch of the Russsian Academy of Sciences, 664033 Irkutsk, Russia
}%

\author{Alexander~Yu.~Sokolov}
\affiliation{
Department of Chemistry and Biochemistry, The Ohio State University, Columbus, Ohio 43210, USA
}%

\author{Achintya Kumar Dutta}

\affiliation{
Department of Chemistry, Indian Institute of Technology Bombay, Mumbai 400076, India
}%
 \email{achintya@chem.iitb.ac.in}
\date{\today}

\title[An \textsf{achemso} demo]
  {Low-Cost Third-order Algebraic Diagrammatic Construction Method for Excited States Using Tensor Hypercontraction and State-Specific Frozen Natural Orbitals}

\abbreviations{IR, NMR, UV}

\begin{document}

\newpage

\begin{abstract}
We have developed a reduced-cost framework for excited-state calculations by combining the state-specific frozen natural orbital (SS-FNO) approach with the third-order algebraic diagrammatic construction [ADC(3)] method and tensor hypercontraction (THC). This strategy enables a systematic and controllable truncation of the virtual orbital space, while significantly lowering the computational scaling through the THC representation of the electron repulsion integrals. The method allows black-box application via adjustable thresholds and includes a perturbative correction that compensates for the truncation error. We have used the CIS(D) method to generate the natural orbitals for the excited states. The performance of the SS-FNO-THC-ADC(3) method is assessed for valence, Rydberg, and charge-transfer excitations. Benchmark calculations demonstrate that the combined approach achieves excellent agreement with the canonical ADC(3) results. The efficiency of the present implementation is demonstrated by calculating the excitation energy of a molecule containing 31 atoms and more than one thousand basis functions.
\end{abstract}
\maketitle

\newpage

\section{Introduction:}
Reliable excited-state theoretical description of the photochemistry of large molecular systems remains a central challenge in electronic structure theory.\cite{gonzalez2012progress, dreuw2005single, plasser2012electronically, knepp2025excited, herbert2016beyond}
Practical calculations for larger systems require a careful balance between two key aspects: accuracy and computational cost. On one hand, excited states are inherently more complex than ground states. They are often energetically close to one another, yet differ significantly in their electronic character. For instance, valence excitations of the $\pi \rightarrow \pi^*$ type may coexist with charge-transfer, Rydberg, or doubly excited states. Since these different types of excitations require distinct theoretical treatments, it is challenging for a single method to describe all of them accurately. Consequently, most computational approaches are tailored to target the specific classes of excitations that are most relevant to the photochemical processes of interest. On the other hand, accurate excited-state methods are computationally demanding, whereas more efficient approaches rely on approximations that may compromise accuracy. A variety of quantum chemical methods are available in the literature, ranging from semi-empirical schemes \cite{ridley1973intermediate, li1999class, voityuk2014indo} through time-dependent density functional theory (TDDFT)\cite{runge1984density, casida1995recent, dreuw2005single} to state-of-the-art wave-function-based methods.\cite{sekino1984linear, stanton1993equation, koch1990coupled} TDDFT is the most popular owing to its low computational cost, but it often fails for conjugated systems and charge-transfer states.\cite{dreuw2005single} Wave-function-based methods are largely free from these problems.\\
Among single-reference wave-function-based methods, coupled-cluster theory shows good accuracy. Polarization propagator-based approaches have also gained attention, mainly the algebraic diagrammatic construction (ADC) based methods. The foundation was laid in 1982 by Schirmer \textit{et al.}\cite{schirmer1982beyond,ISR} with the introduction of the second-order [ADC(2)] and third-order [ADC(3)]\cite{trofimov1999consistent} schemes, which give an average error in excitation energy of 0.2-0.3 eV.\cite{dreuwAlgebraicDiagrammaticConstruction2015,10.1063/1.1504708} 
The extension to fourth order [ADC(4)] has also been achieved.\cite{leitner2022fourth} The computational scaling of excitation energy calculations in ADC(3) is comparable to that of EOM-CCSD, both exhibiting an $N^6$ dependence. However, ADC-based methods offer important advantages, as they are inherently size-consistent and computed properties are size-intensive and rely on a Hermitian formulation. 
Nevertheless, the applicability of ADC(3) is limited to relatively small or medium-sized systems. In particular, the formal $\mathcal{O}(N^6)$ scaling of ADC(3) arises primarily from tensor contractions involving four-index two-electron integrals, specifically those with four virtual indices (VVVV), contracted with the ground-state doubles amplitudes or the doubles components of the excited-state vectors. In systems with large correlation spaces, these contractions also impose significant memory demands, which can become a limiting factor in practical calculations.\\
To overcome these limitations, substantial effort has been devoted in recent years to the development of low-scaling and reduced-cost excited-state methodologies.\cite{duttaPairNaturalOrbital2016,helmichPairNaturalOrbital2013,frankPairNaturalOrbital2018,frank2020implementation,berraud2019unveiling,mester2018reduced,mester2023reduced,manna2025reducedcostequationmotion,manna2024efficient,mukhopadhyay2023state,mukhopadhyay2025reduced,mukhopadhyay2026reducedcostthirdorderalgebraicdiagrammatic} Among the most widely adopted strategies is the density-fitting (DF) approximation,\cite{weigend1998ri,whitten1973coulombic,feyereisen1993use,neese2003improvement,jung2005auxiliary,beebe1977simplifications,aquilante2007unbiased} which replaces four-center two electron repulsion integrals with products of three-center two-electron integrals. The DF-based implementations of ADC\cite{banerjeeEfficientImplementationSinglereference2021,10.1063/1.3048877,10.1063/5.0002851,dempwolff2019efficient,10.1021/acs.jctc.6c00127} significantly reduce memory demands and lower the prefactors of the ``Coulomb-type'' contractions. However, the formal scaling of the ``exchange-like'' terms is not reduced by density fitting alone, which limits its effectiveness for higher-order methods and large basis sets unless special techniques are employed.\cite{10.1063/5.0058766,10.1063/1.4768241} More recently, tensor factorization techniques have emerged as powerful tools for reducing the scaling of correlated electronic structure methods. Among these, tensor hypercontraction (THC)\cite{hohensteinTensorHypercontractionDensity2012,10.1063/1.4768241} provides a particularly compact representation of the two-electron integrals by factorizing them into products of low-rank tensors defined on real-space or auxiliary grids. In contrast to conventional density fitting, THC simultaneously compresses both Coulomb and exchange contributions, leading to substantial reductions in computational scaling and memory usage. THC has been successfully applied to ground-state correlated methods\cite{hohensteinTensorHypercontractionDensity2012,doi:10.1021/acs.jctc.3c01038} and, more recently, to selected excited-state approaches,\cite{manna2025reducedcostequationmotion} demonstrating its potential as a general framework for low-scaling wave-function theory.\\
An alternative and complementary strategy for reducing computational cost is provided by natural-orbital-space truncation techniques.\cite{barrNatureConfigurationInteractionMethod1970,lowdinQuantumTheoryManyParticle1955} These approaches exploit the rapid decay of virtual orbital contributions to electron correlation, enabling a compact representation of the excitation space with minimal loss of accuracy. Accordingly, a variety of lower-scaling excited-state methods based on natural, localized, and pair natural orbitals have been proposed in the coupled-cluster framework.\cite{duttaPairNaturalOrbital2016,helmichPairNaturalOrbital2013,frankPairNaturalOrbital2018,frank2020implementation,berraud2019unveiling} Within the context of ADC methods, Kállay and co-workers have extended natural-orbital-based approximations to achieve reduced scaling\cite{mester2018reduced,mester2023reduced} for the ADC(2) method. Furthermore, our group has developed a state-specific (SS) FNO-based approach for the calculation of valence ionization potentials\cite{mukhopadhyay2023state} and electron affinities\cite{mukhopadhyay2026reducedcostthirdorderalgebraicdiagrammatic} at the ADC(3) level. However, the ADC(3) method for excited states poses additional challenges. The excited-state variant of ADC(3) scales formally as $O(N^{6})$ for both the reference and the target state, which is more expensive than the IP or EA variants, where the target-state solution scales as $O(N^{5})$. Moreover, the most time-consuming contractions in the EE-ADC(3) method are of the ``exchange'' kind, involving integrals with four virtual orbitals, the formal scaling of which cannot be reduced by the DF approximation. The aim of this paper is to develop a low-cost ADC(3) method for excited states based on density fitting, tensor hypercontraction, and frozen natural orbitals.

\section{Theory}

\subsection{Algebraic Diagrammatic Construction Theory}
The algebraic diagrammatic construction (ADC) approach constitutes a hierarchical and systematically improvable framework within many-body propagator theory for computing generalized electronic excitation energies.\cite{schirmer1982beyond,dreuw2015algebraic,schirmer2018many,schirmer1991closed,oddershede1978polarization,ODDERSHEDE198433,oddershede1987propagator,schirmer1983new,ortiz2013electron} The physical observables accessible through ADC depend critically on the choice of the underlying Green’s function. When the formalism is built upon the polarization propagator, ADC yields electronic excitation spectra and optical response properties.\cite{schirmer1982beyond,dreuw2015algebraic,schirmer2018many,trofimov1995efficient,trofimov1997polarization,trofimov1997polarizationII} In contrast, ADC formulations based on the one-particle (electron) propagator provide direct access to charged excitation processes, enabling accurate predictions of ionization potentials and electron affinities.\cite{schirmer2018many,schirmer1983new,von1984computational,schirmer1989green,schirmer1998non,trofimov2005molecular,10.1021/acs.jctc.3c00251}

In its original formulation, the ADC approach did not yield explicit representations of excited-state wave functions. Instead, physical information was obtained indirectly: excitation energies emerged as the pole positions of the relevant propagator, while transition strengths were extracted from the corresponding residues. This shortcoming was subsequently overcome by the introduction of the intermediate-state representation (ISR) formalism.\cite{mertins1996algebraic,schirmer2004intermediate,schirmer2018many,schirmer1991closed,mertins1996algebraicII}

Within the ISR framework, the construction of the excitation space is based on a set of correlated excited states (CES) defined as
\begin{equation}
|\Psi_{J}^{0} \rangle = \hat{C}_{J}|\Psi_{0} \rangle ,
\label{eq:ces}
\end{equation}
where $|\Psi_0\rangle$ denotes the correlated ground state and $\hat{C}_J$ represents an excitation operator acting on it. The explicit form of $\hat{C}_J$ depends on the choice of propagator and, consequently, on the physical process under consideration. For the polarization propagator, the excitation operators generate neutral particle–hole configurations associated with electronic excitation energies (EE). In contrast, formulations based on the one-particle electron propagator involve operators that remove or attach an electron, corresponding to ionization potentials (IP) and electron affinities (EA), respectively:
\begin{align}
\mathrm{EE}\!: \hspace{1cm}\hat{C}_{J} &= \left\{\hat{a}_a^{\dagger}\hat{a}_i;\; \hat{a}_a^{\dagger}\hat{a}_b^{\dagger}\hat{a}_j\hat{a}_i;\; a < b,\; i<j;\; \ldots\right\} \\
\mathrm{IP}\!: \hspace{1cm}\hat{C}_{J} &= \left\{\hat{a}_i;\; \hat{a}_a^{\dagger}\hat{a}_j\hat{a}_i;\; i<j;\; \ldots\right\}\\
\mathrm{EA}\!: \hspace{1cm}\hat{C}_{J} &= \left\{\hat{a}_a^{\dagger};\; \hat{a}_a^{\dagger}\hat{a}_b^{\dagger}\hat{a}_j;\; a < b;\; \ldots\right\}
\end{align}
Here, indices $i,j,\ldots$ label occupied orbitals, $a,b,\ldots$ denote virtual orbitals, and $p,q,\ldots$ represent general orbital indices. The excitation operators can be systematically categorized according to their $\mu$-hole–$\mu$-particle ($\mu h$–$\mu p$) character, where $\mu$ counts the number of annihilation and creation operators involved. An analogous classification scheme applies in the IP and EA sectors. Each excitation operator $\hat{C}_J$ is assigned to a class $[J]$, with the correlated ground state corresponding to the zeroth class, $[J]=0$.

The correlated excited states defined in Eq.~\eqref{eq:ces} do not, in general, constitute an orthogonal set.\cite{mertins1996algebraic} To remedy this, the ISR framework employs a hierarchical two-stage orthogonalization procedure.\cite{schirmer1991closed} In the first stage, a class-wise Gram–Schmidt (GS) orthogonalization is applied to the correlated excited states, yielding a set of so-called precursor states.
\begin{equation}
|\Psi_J^\#\rangle = |\Psi_J^0\rangle - \sum_{K\,[K]<[J]} |\tilde{\Psi}_K\rangle \langle \tilde{\Psi}_K | \Psi_J^0 \rangle ,
\end{equation}
In the second stage, the precursor states within the same excitation class are transformed into an orthonormal set by means of symmetric orthogonalization,
\begin{equation}
|\tilde{\Psi}_I\rangle = \sum_{J\,[J]=[I]} |\Psi_J^\#\rangle S_{IJ}^{-1/2},
\end{equation}
where the overlap matrix is defined as
\begin{equation}
    S_{IJ} = \langle \Psi_I^\# | \Psi_J^\# \rangle.
\end{equation}

The resulting set of intermediate states ${|\tilde{\Psi}_I\rangle}$ forms a complete orthonormal basis, which serves as the working representation for the construction of the ISR secular (ADC) matrix $\mathbf{M}$.
\begin{equation}
M_{IJ} = \langle \tilde{\Psi}_{I}| \hat{H}-E_0|\tilde{\Psi}_{J}\rangle ,
\end{equation}
which represents the shifted Hamiltonian $\hat{H}-E_0$ in this intermediate-state basis. The exact $N$-electron eigenstates $|\Psi_n^{N}\rangle$ are represented within the ISR framework as linear combinations of the orthonormal intermediate states,
\begin{equation}
|\Psi_n^{N}\rangle = \sum_{I} X_{In} |\tilde{\Psi}_I\rangle ,
\end{equation}
where the expansion coefficients $X_{In}$ are obtained from the eigenvectors of the ISR secular matrix $\mathbf{M}$. In a compact matrix formulation, the associated eigenvalue problem reads
\begin{equation}
\mathbf{M}\mathbf{X} = \mathbf{X}\boldsymbol{\Omega},
\qquad
\mathbf{X}^\dagger \mathbf{X} = \mathbf{1},
\label{eq10}
\end{equation}
with the diagonal matrix $\boldsymbol{\Omega}$ collecting the excitation energies.

The secular matrix $\mathbf{M}$ admits a systematic expansion in orders of the perturbation operator. Truncation of this expansion at a specified order $n$ yields the working equations of the ADC($n$) hierarchy,
\begin{equation}
\mathbf{M}
= \mathbf{M}^{(0)} + \mathbf{M}^{(1)} + \mathbf{M}^{(2)} + \mathbf{M}^{(3)} + \cdots {.}
\label{eq11}
\end{equation}
Restricting $n$ to 3 leads to the ADC(3) method. An alternative derivation of the ADC(3) method is possible\cite{10.1063/1.5055380} using the effective Liouvillian formalism of Mukherjee and Kutzelnigg.\cite{mukherjeeEffectiveLiouvilleanFormalism1989}
Almost all EE-ADC(3) implementations now use a spin-orbital approach; here, we present an RHF-based implementation for a closed-shell ground state reference and $\hat{S}^2$- and $\hat{S}_z$-adapted intermediate states describing singlet excitations. The triplet excited states and transition properties are not currently available for the spin-adapted RHF reference; however, they can be calculated using the spin-orbital-based implementation\cite{10.1021/acs.jctc.6c00127} in BAGH.\cite{dutta2025bagh}

\subsection{Density Fitting}

The density fitting approximation\cite{weigend1998ri,whitten1973coulombic,feyereisen1993use,neese2003improvement,jung2005auxiliary,beebe1977simplifications,aquilante2007unbiased} can greatly speed up the calculation of two-electron integrals over atomic basis functions and their subsequent manipulation. The four-center two-electron integral can be expressed in terms of orbital densities as
\begin{equation}
(p q \mid r s)=\int d\mathbf{r}_1 \int d\mathbf{r}_2\; \rho_{pq}(\mathbf{r}_1)\frac{1}{r_{12}}\rho_{rs}(\mathbf{r}_2)
\label{eq:19}
\end{equation}
where $\rho_{pq}(\mathbf{r})=\varphi_{p}(\mathbf{r})\varphi_{q}(\mathbf{r})$. Each density can then be approximated in terms of auxiliary basis functions $\chi_{P}(\mathbf{r})$ as
\begin{equation}
\tilde{\rho}_{pq}(\mathbf{r}) = \sum_{P}^{N_{aux}}d_{P}^{pq}\chi_{P}(\mathbf{r}),
\label{eq:20}
\end{equation}
where the fitting coefficient $d_P^{pq}$ is obtained by minimizing the following error functional
\begin{equation}
\xi_2 = \frac{1}{2}
\int \int
\frac{\left[\rho(\mathbf{r}_1) - \tilde{\rho}(\mathbf{r}_1)\right]
      \left[\rho(\mathbf{r}_2) - \tilde{\rho}(\mathbf{r}_2)\right]}
     {\left|\mathbf{r}_1 - \mathbf{r}_2\right|}
\, d\mathbf{r}_1 \, d\mathbf{r}_2 .
\end{equation}
This leads to the fitting coefficient
\begin{equation}
d_P^{pq} = \sum_{Q}(pq|Q)\,[J^{-1}]_{QP}
\label{eq:21}
\end{equation}
where
\begin{equation}
(pq|Q) = \int d\mathbf{r}_1 \int d\mathbf{r}_2\; \varphi_{p}(\mathbf{r}_1)\varphi_{q}(\mathbf{r}_1)\frac{1}{r_{12}}\chi_{Q}(\mathbf{r}_2)
\label{eq:22}
\end{equation}
and the metric integral
\begin{equation}
J_{PQ} = \int d\mathbf{r}_1 \int d\mathbf{r}_2\; \chi_{P}(\mathbf{r}_1)\frac{1}{r_{12}}\chi_{Q}(\mathbf{r}_2).
\label{eq:23}
\end{equation}
Using the resolution-of-the-identity (RI) approximation, the four-center two-electron repulsion integrals (ERIs) can be written as
\begin{equation}
 (p q \mid r s)=\sum_{P Q}(p q \mid P)(P \mid Q)^{-1}(Q \mid r s)
\end{equation}
In practice, the inverse matrix is factorized into the product of its inverse-square-root matrices, and the four-center ERIs are recast as
\begin{equation}
(p q \mid r s)=\sum_Q J_{p q}^Q J_{r s}^Q
\end{equation}
where
\begin{equation}
J_{p q}^Q=\sum_P(p q \mid P)(P \mid Q)^{-\frac{1}{2}}
\end{equation}
All integrals in the current implementation are computed using the RI approximation. Integrals with up to two virtual indices are constructed and stored in memory or on disk. Integrals with three or four virtual indices are not stored; the terms involving the contraction of the trial excited-state vector amplitudes ($\mathbf{r}$) with these integrals are factorized in terms of the three-center two-electron integral $J_{p q}^Q$. The explicit equations for the factorized terms are provided in the supplementary material.

\subsection{Natural auxiliary function}
\label{sec2.4}
The auxiliary basis set can be reduced using natural auxiliary functions (NAFs), introduced by K\'allay and co-workers,\cite{kallaySystematicWayCost2014} in the same way that natural orbitals are used to truncate the orbital basis set. In the NAF approach, the three-center integral matrix $J$ (with the auxiliary index as its row index) is written by singular value decomposition (SVD) as
\begin{equation}
J = M \Sigma N^{T}
\end{equation}
Here, $M$ and $N$ contain the left and right singular vectors, respectively, and the diagonal matrix $\Sigma$ holds the singular values. In practice, the singular values and left singular vectors of $J$ are obtained by diagonalizing the symmetric matrix
\begin{equation}
    W = J J^{T}.
\end{equation}
Its eigenvalues are equal to the squares of the singular values, and its eigenvectors are the left singular vectors of $J$. Here, $W$ is an $n_{\text{aux}} \times n_{\text{aux}}$ matrix, where $n_{\text{aux}}$ is the size of the auxiliary basis set. The NAFs with eigenvalues lower than a threshold $(\varepsilon_{NAF})$ are dropped, and the remaining $\overline{n}_{aux}$ eigenvectors are collected in the ${n}_{\text{aux}} \times \bar{n}_{\text{aux}}$ matrix $\overline{M}$. The auxiliary-function index of $J$ is then transformed by $\overline{M}$ as
\begin{equation}
    \overline{J}= \overline{M}^{T} J
\end{equation}
The resulting matrix $\overline{J}$ provides the best rank-$\overline{n}_{aux}$ representation of $J$ in the least-squares sense, so that $(pq|rs) \approx \sum_{\bar{Q}} \overline{J}_{pq}^{\,\bar{Q}}\,\overline{J}_{rs}^{\,\bar{Q}}$. Consequently, the reduction in computational time in the NAF basis is expected to be proportional to the ratio $\frac{n_{aux}}{\overline{n}_{aux}}$. One can use any of the three-center two-electron integrals $J_{ij}^{Q}$, $J_{ia}^{Q}$, or $J_{ab}^{Q}$ to generate natural auxiliary functions. Following the suggestion of K\'allay and co-workers,\cite{mester2017reduced} we use $J_{ab}^{Q}$ to generate the natural auxiliary functions.

\subsection{Frozen Natural Orbitals}
\label{sec2.5} 

Natural orbitals (NOs)\cite{lowdinQuantumTheoryManyParticle1955} are generated by diagonalizing the one-body reduced correlated density matrix. Among the various flavors of natural orbitals available, we use the frozen natural orbital\cite{barrNatureConfigurationInteractionMethod1970} (FNO) scheme, in which the occupied orbitals are kept frozen at the HF level and the virtual orbitals are expanded in terms of natural orbitals. The MP2 method is generally used to construct the one-body reduced density matrix for the FNOs. Its virtual-virtual block is given by
\begin{equation}
\label{eq:density}
D_{ab}^{\text{MP2}} = \sum_{ijc} \left( 2t_{ij}^{ac}t_{ij}^{bc} - t_{ij}^{ac}t_{ij}^{cb} \right)
\end{equation}
where $t_{ij}^{ab}$ is an MP2 doubles amplitude defined as
\begin{equation}
t_{ij}^{ab} = \frac{(ai|bj)}{\varepsilon_i + \varepsilon_j - \varepsilon_a - \varepsilon_b}
\end{equation}
and the $\varepsilon$ are the Hartree--Fock orbital energies. Diagonalizing the density matrix of Eq.~\eqref{eq:density} yields the FNOs as the eigenvectors ($V$) and the corresponding occupation numbers $n$,
\begin{equation}
D_{vv}^{\text{MP2}}V= Vn
\end{equation}
FNOs with small occupation numbers make a negligible contribution to the correlation energy, and the corresponding virtual orbitals can be omitted from the correlation calculation. One therefore chooses a predefined threshold below which natural orbitals are removed from the virtual space. The virtual-virtual block of the Fock matrix is then transformed into the truncated natural orbital basis,
\begin{equation}
\label{eq:fnofock}
\widetilde{F}_{vv} = \widetilde{V}^{\dagger} F_{vv} \widetilde{V}
\end{equation}
where ${F}_{vv}$ is the virtual--virtual block of the Fock matrix and $\widetilde{V}$ collects the retained natural virtual orbitals. Diagonalizing this Fock matrix gives the semi-canonical orbital energies $\widetilde{\varepsilon}$ in the truncated natural orbital basis,
\begin{equation}
\label{eq:fnodiag}
\widetilde{F}_{vv}\widetilde{Z}= \widetilde{Z}\widetilde{\varepsilon}
\end{equation}
The matrices
\begin{align}
\widetilde{U}_{vir} &= U_{vir}\widetilde{V}\widetilde{Z} \\
\widetilde{U}_{occ} &= U_{occ}
\label{eq:fnotrans}
\end{align}
define the transformation from the canonical MO basis to the semi-canonical frozen natural orbital basis for the virtual and occupied spaces, respectively. The molecular orbital coefficients are represented by the matrix ${U}$, whose rows correspond to atomic orbitals and columns to molecular orbitals. The occupied and virtual orbital coefficient matrices are denoted by ${U}_{\mathrm{occ}}$ and ${U}_{\mathrm{vir}}$, respectively.

\subsection{State-Specific Frozen Natural Orbitals}
Generating frozen natural orbitals from the MP2 density is the standard choice in FNO-based calculations. Natural orbitals obtained this way are optimized for recovering the ground-state correlation energy but are not well suited for the calculation of excitation energies\cite{manna2025reducedcostequationmotion,mukhopadhyay2025reducedEA} or response properties.\cite{10.1021/acs.jpca.6b11410} There is no consensus on how to generate an appropriate set of natural orbitals for excited states. Helmich and H\"attig\cite{helmich2011local} used state-specific pair natural orbitals (PNOs) obtained from the CIS(D) method to generate an approximate virtual space for the excited states. K\'allay and co-workers,\cite{mester2017reduced} on the other hand, used state-specific excited-state natural orbitals generated from the CIS(D) density to reduce the computational cost of CC2 calculations. Valeev and co-workers\cite{peng2018state} used state-averaged PNOs for excited states. Neese and co-workers\cite{dutta2018exploring} took an alternative strategy, combining a similarity transformation with the ground-state PNOs to arrive at a low-cost coupled-cluster-based excited-state method. We have recently shown that the ADC(2) method gives an accurate first-order description of the excited-state wave function, and that an ADC(2)-natural-orbital-based EOM-CCSD method gives uniform accuracy for valence, Rydberg, and charge-transfer states. \cite{manna2025reducedcostequationmotion}
However, frozen natural orbitals generated from the state-specific CIS(D) density give an accuracy almost comparable to that of the ADC(2) orbitals, with a significant reduction in the cost of natural orbital generation.\cite{manna2025reducedcostequationmotion} For the $k$th excited state, the virtual-virtual block of the state-specific density matrix is obtained as
\begin{equation}
\label{eq:ssdens}
D_{ab}^{SS}(k) = D_{ab}^{MP2}+D_{ab}^{CIS(D)}(k)
\end{equation}
where $D_{ab}^{CIS(D)}(k)$ is the CIS(D) excited-state density matrix for the $k$th state,
\begin{equation}
D_{ab}^{CIS(D)}(k)=\sum_{i}c_{i}^{a}c_{i}^{b}+\sum_{ijc}(2c_{ij}^{ca}c_{ij}^{bc}-c_{ij}^{ca}c_{ij}^{cb})
\end{equation}
and $c_{i}^{a}$ and $c_{ij}^{ab}$ are the CIS(D) singles and doubles vectors.
The state-specific natural orbitals obtained from Eq.~\eqref{eq:ssdens} are substituted into Eqs.~\eqref{eq:fnofock}--\eqref{eq:fnotrans}. The subsequent CIS(D) and ADC(3) calculations are performed in the state-specific frozen natural orbital (SS-FNO) basis.

\subsection{State-Averaged Frozen Natural Orbitals}
As in the state-specific approach, the correlated ground-state (MP2) density $D_{ab}^{MP2}$ is constructed once. This contribution is state-independent and is the same for all excited states. The CIS(D) excited-state density is evaluated for each of the selected states and averaged over them; the resulting mean density is added to the ground-state density to form the total state-averaged density used for the excited-state computation. The virtual-virtual block of the state-averaged density matrix is
\begin{equation}
\label{eq:sadens}
D_{ab}^{SA} = D_{ab}^{MP2}+ \frac{1}{N_{\text{roots}}}\sum_{k=1}^{N_{\text{roots}}}D_{ab}^{CIS(D)}(k)
\end{equation}
The state-averaged density matrix of Eq.~\eqref{eq:sadens} is inserted into Eqs.~\eqref{eq:fnofock}--\eqref{eq:fnotrans} to obtain the transformation matrix for the state-averaged frozen natural orbital basis; this transformation is performed only once and is then used for all excited states. When only a limited number of roots is calculated, the state-averaged natural orbitals offer an advantage over the state-specific approach, for which the excited-state density, integral transformation, and ground-state calculation must be repeated for every state. However, as the number of excited states increases, the compression offered by the state-averaged approach becomes minimal.

\subsection{Correction for Truncation Errors}
Perturbative corrections for natural orbital truncation have been shown to significantly improve the accuracy of natural-orbital-based ground-state coupled-cluster calculations. We use the CIS(D) method, which also generates the SS/SA-FNOs, to estimate the error introduced by natural orbital truncation. The corrected SS-FNO-EE-ADC(3) excitation energies are obtained by adding the difference between the canonical and SS-FNO CIS(D) excitation energies to the SS-FNO-EE-ADC(3) result:
\begin{align}
\omega_{\mathrm{EE\text{-}ADC(3)}}^{\mathrm{Canonical}}
&\approx
\omega_{\mathrm{SS\text{-}FNO\text{-}EE\text{-}ADC(3)}}^{\mathrm{Corrected}}
\nonumber\\
&=
\omega_{\mathrm{SS\text{-}FNO\text{-}EE\text{-}ADC(3)}}^{\mathrm{Uncorrected}}
+\left(
\omega_{\mathrm{CIS(D)}}^{\mathrm{Canonical}}
-
\omega_{\mathrm{CIS(D)}}^{\mathrm{SS\text{-}FNO}}
\right).
\end{align}
\subsection{Tensor Hypercontraction Implementation of the Four-External Term}
TThe most time-consuming term in the EE-ADC(3) equations is the contraction of the four-virtual two-electron integral, also referred to as external integrals, with the doubles amplitudes, which occurs in both the ground- and excited-state equations and is of the exchange type,
\begin{equation}
W_{ij}^{ab} = \sum_{cd} \left(1+P(ij)P(ab)\right)(ac|bd)\, r_{ij}^{cd}
\end{equation}
Here, $P$ denotes the permutation operator. The leading cost of evaluating this term scales as $\mathcal{O}(N_{occ}^{2}N_{vir}^{4})$, or $\mathcal{O}(N_{vir}^{4}N_{aux})$ in a density-fitted implementation. Although density fitting cannot reduce the formal scaling of the exchange-type terms in the EE-ADC(3) method, the tensor hypercontraction (THC) approximation\cite{10.1063/1.4768241} lowers the overall scaling by factorizing the integrals into lower-order tensors. In THC, the four-center two-electron integrals are written as a product of five matrices,
\begin{equation}
\label{eq:thc}
    \left( pq\text{ }\!\!|\!\!\text{ }rs \right)\approx {{\left( pq\text{ }\!\!|\!\!\text{ }rs \right)}_{THC}}=\sum\limits_{RS}{X_{p}^{R}X_{q}^{R}{{Z}_{RS}}X_{r}^{S}X_{s}^{S}}
\end{equation}
The grid-based least-squares THC (LS-THC) method\cite{hohensteinTensorHypercontractionDensity2012,10.1063/1.4768241} is the standard approach for THC construction. There, the collocation matrix $X$ is predetermined by evaluating the spatial MOs on a grid,
\begin{equation}
 X_{p}^{R}={{\phi }_{p}}\left( {{r}_{R}} \right)\left( {{w}_{R}} \right)^{1/4}
\end{equation}
where $\mathbf{r}_R$ and $w_R$ denote the grid points and their weights. In LS-THC, the decomposition in Eq.~{\eqref{eq:thc} is constructed through an implicit real-space quadrature by minimizing a cost function defined as the Frobenius norm of the error between the exact and approximated electron repulsion integrals,
\begin{equation}
O=
\frac{1}{2} \left\| (pq | rs)
- \sum_{{RS}} X_{p}^{R} X_{q}^{R} Z_{RS} X_{r}^{S} X_{s}^{S}
\right\|_F^2
\end{equation}
After constructing the $X$ matrices, the $Z$ matrix is obtained by minimizing the cost function $O$,
\begin{align}
  & {{Z}_{RS}}=\sum\limits_{{R}'{S}'}{{{\left( {{S}^{-1}} \right)}_{{R}'R}}{{E}_{{R}'{S}'}}{{\left( {{S}^{-1}} \right)}_{{S}'S}}} 
\end{align}
where
\begin{align}
  & {{E}_{RS}}=\sum\limits_{pqrs}{X_{p}^{R}X_{q}^{R}{{\left( pq\text{ }\!\!|\!\!\text{ }rs \right)}_{DF}}X_{r}^{S}X_{s}^{S}} \\ 
 & \ \quad \ \,=\sum\limits_{Q}{\left( \sum\limits_{pq}{X_{p}^{R}X_{q}^{R}J_{pq}^{Q}} \right)}\left( \sum\limits_{rs}{X_{r}^{S}X_{s}^{S}J_{rs}^{Q}} \right) \\ 
 & \quad \ \,\ =\sum\limits_{Q}{{{\zeta }^{RQ}}{{\zeta }^{SQ}}} 
\end{align}
and the grid metric matrix $S$ is defined as
\begin{equation}
    {{S}_{RS}}=\sum\limits_{pq}{X_{p}^{R}X_{q}^{R}X_{p}^{S}X_{q}^{S}}
\end{equation}
After evaluating the atomic orbital (AO) values on the real-space grid, the electron density and its derivatives are computed by contracting the AO values with the density matrix, as implemented in the PySCF DFT quadrature framework.\cite{sun2018pyscf,972db55628fa4100a640cce8edebe588} 
To reduce the computational cost, grid points with negligible electron density are removed using a density-based pruning scheme.\cite{sun2018pyscf} The AO values are then re-evaluated on the pruned grid and transformed to the molecular orbital (MO) basis.
Once the collocation matrix $X$ and the $Z$ matrix have been generated, they are applied to the most time-consuming step of ADC(3), in which the doubles amplitudes $r_2$ are contracted with the four-external (VVVV) integrals. Using the THC approximation, this contraction becomes
\begin{equation}
W_{ij}^{ab} = \sum_S X_a^S
\left(\sum_R X_b^R {Z_{RS}}\left(
\sum_d X_d^R\left(\sum_c X_c^S r_{ij}^{cd}\right)\right)\right)
\end{equation}
This contraction scales as $\mathcal{O}(N^2_{\mathrm{grid}} N_{\mathrm{vir}} N_{\mathrm{occ}}^2)$, a formal reduction relative to the canonical term. The THC approach has the advantage over atomic-orbital pseudo-spectral methods such as the chain-of-spheres (COSX) approximation that it can be applied directly in the molecular orbital basis, so that it combines naturally with the natural-orbital truncation framework. Matthews and co-workers\cite{10.1021/acs.jctc.3c00892} reported a THC-based acceleration of the particle-particle ladder term in EOM-CCSD; THC is even more advantageous in ADC(3), since there it acts on the bare four-external integrals rather than their ``dressed'' EOM-CCSD counterpart.

\subsection{General Algorithm}
The ADC(3) equations are solved using the iterative Davidson diagonalization procedure.\cite{davidsonIterativeCalculationFew1975} A root-wise Davidson solver is used for the SS-FNO-ADC(3) equations. The CIS(D) vectors used to generate the state-specific densities are transformed to the corresponding SS-FNO basis and used as the guess vector for that root in SS-FNO-ADC(3). The overlap with the guess vector is monitored during the iterations to ensure correct root following and to avoid root flipping. An overview of the SS-FNO-EE-ADC(3) algorithm is as follows:
\begin{enumerate}
    \item Solve the Hartree--Fock (HF) equations.

    \item Generate the three-center two-electron integrals using the density fitting (DF) approximation.

    \item Generate the THC $X$ and $Z$ matrices.

    \item Solve the MP2, CIS, and CIS(D) equations for all excited states in the canonical basis.

    \item Compute the ground-state (MP2) density.
    \item Loop over the excited states obtained from CIS(D):
    \begin{enumerate}
    \item Construct the excited-state density matrix from Eq.~{\eqref{eq:ssdens}}.
    \item Generate the excited-state FNOs and the MO-to-FNO transformation matrix.
    \item Transform the virtual MO indices of $J$ to the excited-state FNO basis $\tilde{J}$.
    \item {Generate $W$ from $\tilde{J}$ and transform the auxiliary index of $\tilde{J}$ to the NAF basis}.
    \item {Transform the THC matrices to the SS-FNO basis.}
    \item Perform {the} CIS(D) calculation in the truncated SS-FNO basis.
    \item Transform the CIS(D) vector to the SS-FNO basis and use it as the guess.
    \item Perform the EE-ADC(3) calculation in the truncated SS-FNO basis.
    \item Evaluate the perturbative correction to the SS-FNO-EE-ADC(3) excitation energy.

    \end{enumerate}
\end{enumerate}
In the canonical ADC(3) method, the most time-consuming step has an iterative scaling of $\mathcal{O}(N_{\mathrm{aux}} N_{\mathrm{vir}}^3 N_{\mathrm{occ}}^2)$, {where $N_{\mathrm{aux}}$, $N_{\mathrm{vir}}$, and $N_{\mathrm{occ}}$ are the numbers of auxiliary basis functions, virtual orbitals, and occupied orbitals, respectively}.\\The {combined} use of tensor hypercontraction, frozen natural orbitals (FNO){,} and natural auxiliary functions (NAF) reduces the {cost} of this step to $\mathcal{O}(N^2_{\mathrm{grid}} (N_{\mathrm{vir}}-N_{\mathrm{vir}}^{f}) N_{\mathrm{occ}}^2)$, where $N_{\mathrm{vir}}^{f}$ is the number of discarded virtual orbitals and $N_{\mathrm{grid}}$ is the number of pruning grid points. The choice and convergence of $N_{\mathrm{grid}}$ is investigated in Sec. IV A. The construction of FNOs introduces {an additional $\mathcal{O}(N_{\mathrm{vir}}^3 N_{\mathrm{occ}}^2)$ cost}; however, {this} is negligible compared to the total computational time and the THC decomposition introduces additional storage for the factor matrices, this cost is considerably smaller than that of storing the full four-index tensor for the THC ranks used here.

\section{Computational Details}

The SS-FNO-EE-ADC(3) method has been implemented in our in-house software package BAGH,\cite{dutta2025bagh} which is written in Python, with Cython and Fortran {used for} the computationally demanding parts. It is interfaced with PySCF,\cite{sun2015libcint,sun2018pyscf,sun2020recent} DIRAC,\cite{bast2023dirac23} and GAMESS-US.\cite{barca2020recent} All {calculations presented here were performed through the} PySCF interface of BAGH.

\section{Results and Discussion}
\subsection{Benchmarking the THC Grid and Cutoff}
To generate the THC integrals {with} a grid-based approach, a spatial integration grid is first constructed. This grid is significantly larger than {both} the auxiliary and the atomic orbital (AO) {basis}. {A separate benchmarking study is therefore} necessary to determine {an} optimal THC pruning threshold and grid {for} accurate excited-state {calculations}. \\
As a benchmark, the first excitation energy of propanamide is computed {in} the TZVP basis set, for which the reference (canonical) value is 5.81 eV. The geometry of propanamide was taken from {Ref.~\cite{doi:10.1021/acs.jpclett.9b03652}}. First, the radial grid is varied from 1 to 9 while {the} angular grid {is} fixed at 38. This increases the total number of grid points from $(456,137)$ to $(3904,137)$, where 137 {is} the AO basis size of propanamide in the TZVP basis set. {Figure~\hyperref[fig:radial-angular-grid]{1}(a) shows that, as the grid is made denser, the error in the excitation energy decreases from about 0.02~eV to essentially zero (the energy rising from 5.79~eV to the reference value of 5.81~eV)}. Beyond the $(5,38)$ grid, the change in energy is {$\leq 0.01\ \text{eV}$}, {i.e., the result is} essentially converged. {A radial grid of 6 [the $(6,38)$ grid, 2616 points] is therefore sufficient, and the angular grid is examined next}.\\
Next, the radial grid is fixed at 6{,} and the angular grid is varied from 1 to 38, {which changes} the number of grid points from $(72,137)$ to $(2616,137)$. {As shown in Figure~\hyperref[fig:radial-angular-grid]{1}(b), the error in the first excitation energy is already 0.00~eV at the $(6,14)$ grid and changes negligibly beyond $(6,6)$}. {A} THC grid of $(6,14)$ is {therefore used} for all ADC(3) calculations in this work.\\
Finally, the THC pruning threshold is varied from $10^{-1}$ to $10^{-5}$ (see Table~\hyperref[table:pruning]{S III}){ while keeping} the grid fixed at $(6,14)$. {On this grid}, the total number of grid points for propanamide is 1008, which {is reduced} to 720 {by a} pruning threshold of $10^{-1}$. The excitation energy remains unchanged (0.00 eV error) {over} this range of thresholds, {so} even a relatively {loose} threshold is sufficient. {A} pruning threshold of $10^{-1}$ is {therefore} adopted for all subsequent calculations.
 
\begin{figure*}
    \centering
    
    \begin{subfigure}{0.48\textwidth} 
        \includegraphics[width=\linewidth]{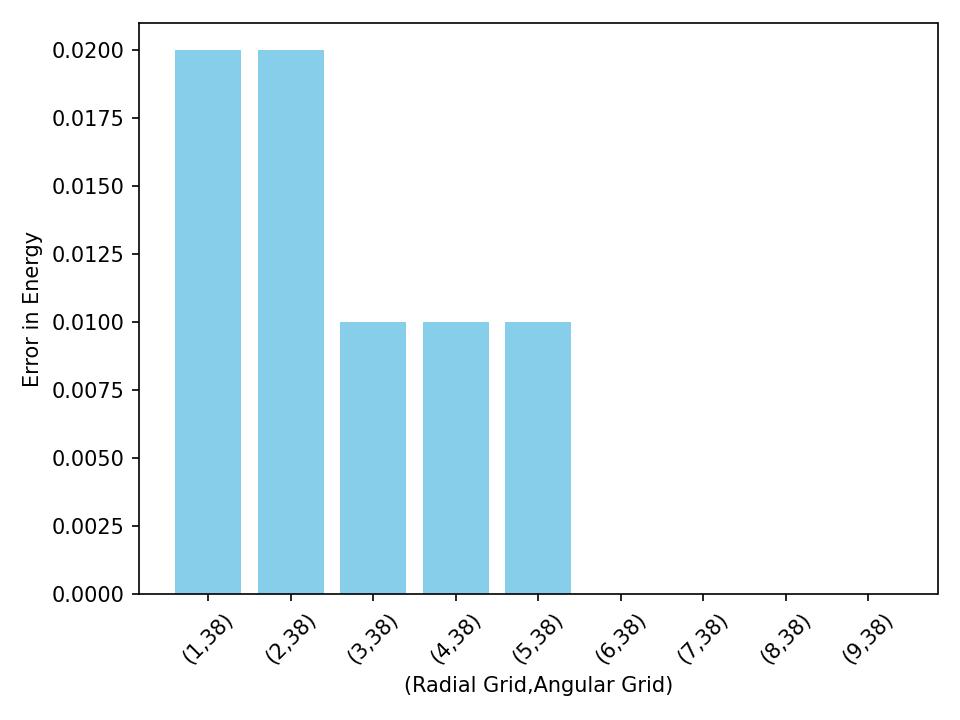} 
        \caption{}
        \label{fig:sub1}
    \end{subfigure}
    \begin{subfigure}{0.48\textwidth}
        \includegraphics[width=\linewidth]{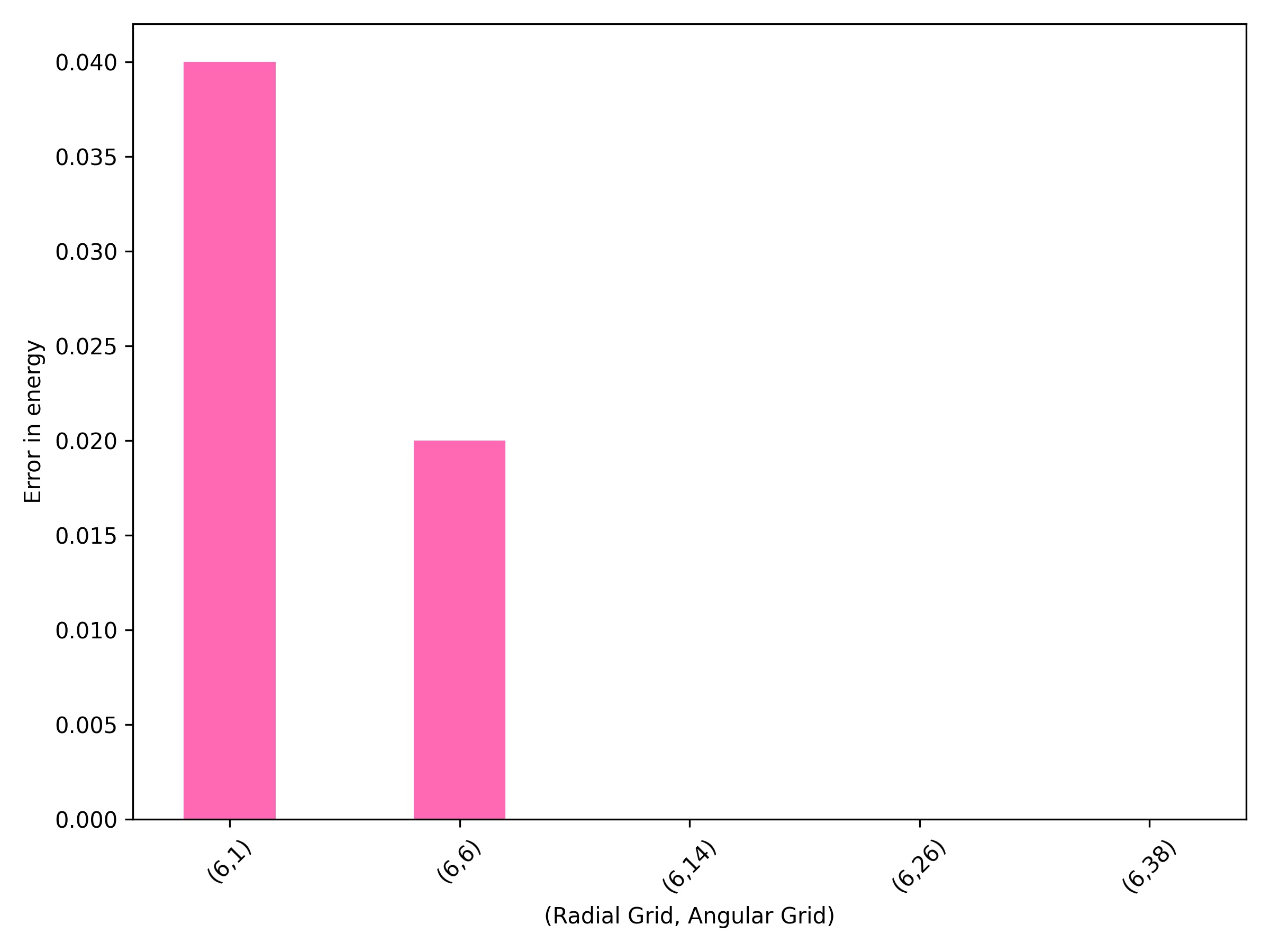}
        \caption{}
        \label{fig:sub2}
    \end{subfigure}
    
    \caption{{Error in the excitation energy of propanamide (TZVP basis set) relative to the canonical value of 5.81~eV. (a) Variation with the radial grid, with the angular grid fixed at 38; (b) variation with the angular grid, with the radial grid fixed at 6.}}
    \label{fig:radial-angular-grid}
\end{figure*}

\subsection{Convergence of FNO threshold}
The accuracy of the SS-FNO-EE-ADC(3) method depends on the FNO truncation threshold {used for} the excited states. Figure~\hyperref[fig:fno_combined]{2}{(a)} presents the convergence of the {first} excitation energy of propanamide with respect to the excited{-state} FNO threshold {in the} TZVP basis set. The {NAF} threshold was set to zero, and canonical EE-ADC(3) with the {full} virtual{space} was used as {the} reference {(total basis set dimension 137)}. The {uncorrected} SS-FNO excitation energy converges systematically with the excited-state truncation threshold {and reaches} the canonical result at a threshold of $10^{-4}$, which {retains about 89\%} of the {virtual} space. The perturbative correction {improves the results substantially: with it, the excitation energy is essentially converged already at a threshold of $10^{-3}$, which retains only 48\% of the virtual space, with an error of 0.01~eV}. {We use a} threshold {of $10^{-4}$} for the excited states {in} all subsequent calculations. {A given threshold may retain a different fraction of the virtual orbitals for different molecules and basis sets; in small basis sets the truncation may be negligible. Nevertheless, the scheme yields consistent accuracy regardless of the fraction of the virtual space retained}.
\begin{figure*}
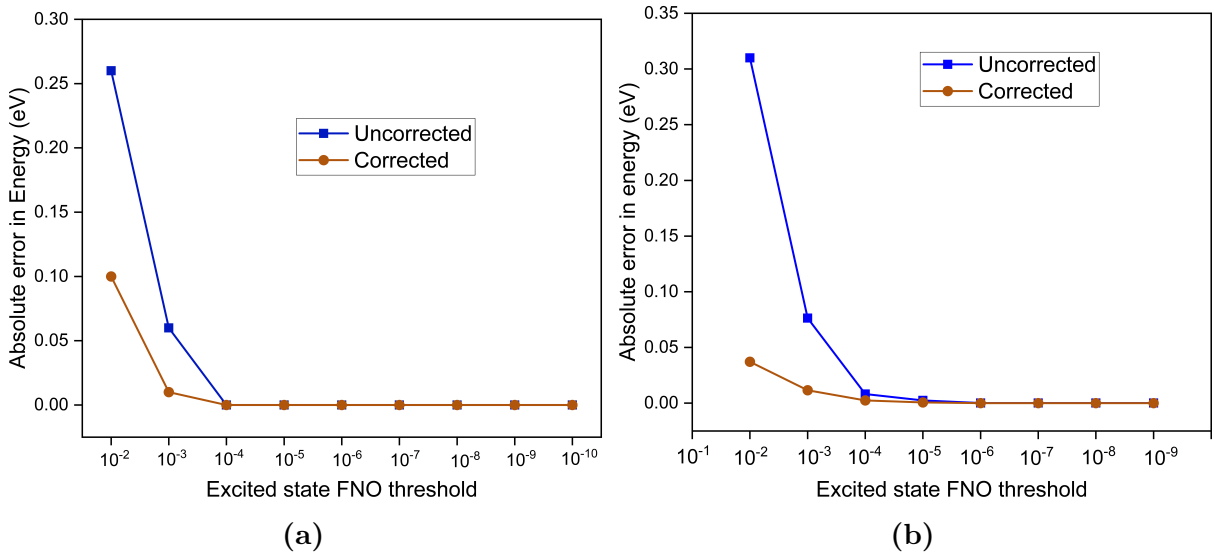

    \centering
    \begin{subfigure}{0.48\textwidth}
        \centering
        \includegraphics[width=\linewidth]{figure/fnothreshold_vary.jpg}
        \caption{}
        \label{fig:fnovariation}
    \end{subfigure}
    \begin{subfigure}{0.48\textwidth}
        \centering
        \includegraphics[width=\linewidth]{figure/propanamide_aug_cc_pvtz.jpg}
        \caption{}
        \label{fig:fnovariation-avtz}
    \end{subfigure}

    \caption{\label{fig:fno_combined}
    Convergence of the error {in the} SS-FNO-EE-ADC(3) {excitation energy} for the valence excited state of propanamide with respect to the excited-state FNO threshold, {in (a) the TZVP basis set and (b) the aug-cc-pVTZ basis set}. The canonical EE-ADC(3) result with 100\% active virtual orbitals is used as the reference.}
\end{figure*}
\\
{Figure~\hyperref[fig:fno_combined]{2}(b) shows the same convergence behavior in the larger aug-cc-pVTZ basis set, confirming that the $10^{-4}$ threshold transfers to larger basis sets.}
To determine the optimal {threshold for the} natural auxiliary functions (NAFs), we {plot} the {lowest} excitation energy of propanamide {computed with} SS-FNO-EE-ADC(3) {for a range of NAF thresholds} in Figure~\hyperref[fig:natauxvariation]{3}. Throughout these calculations, the excited{-state} FNO threshold is fixed at $10^{-4}${, and the TZVP basis set is used}. The auxiliary basis set is {generated automatically by PySCF and comprises 903 functions} for propanamide. {A} NAF {threshold} of $10^{-2}$ leads to an error {below} 0.001 eV {while retaining} only 50\% of the auxiliary {functions}. This {shows that NAFs can effectively} reduce the {cost} of density-fitting-based ADC calculations. The perturbative correction has a negligible effect on the {NAF truncation error}. {A} NAF threshold {of $10^{-2}$ is therefore used} for the excited states {throughout the rest of the manuscript}.

\begin{figure}
    \centering
    \includegraphics[width=0.5\textwidth]{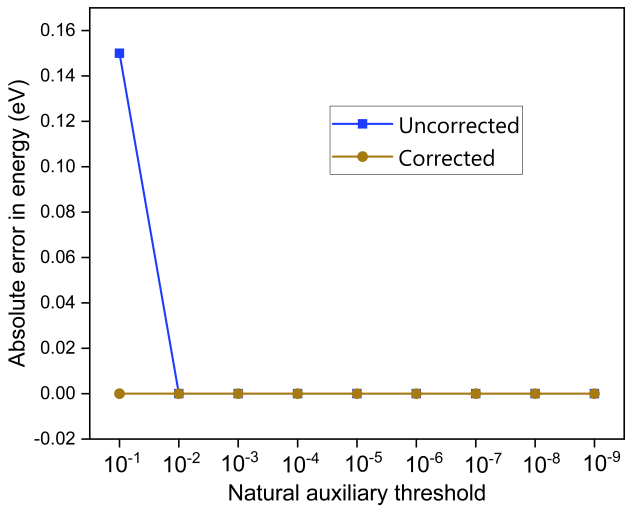}
    \caption{\label{fig:natauxvariation} {Convergence of the error in the SS-FNO-EE-ADC(3) excitation energy of the valence excited state of propanamide with respect to the NAF truncation threshold ($10^{-n}$) in the TZVP basis set}.}
\end{figure}

\subsubsection{Valence Test Set}
{Valence excited states have} been extensively benchmarked{, thanks to} the well-{established} benchmark set {of} Schreiber{, Thiel, and co-workers}.\cite{schreiber2008benchmarks} A statistical analysis of the SS-FNO-EE-ADC(3) results is presented in Table~\hyperref[tab:thiel-tzvp]{I}{, and the individual excitation energies are given in the supplementary material}. The TZVP basis set {is used, and untruncated canonical EE-ADC(3) results serve as the reference}. The maximum absolute error {of} SS-FNO-EE-ADC(3) with {CIS(D) SS-FNOs} is 0.029 eV{; the perturbative correction reduces it marginally, to} 0.026 eV. The RMSD {of} SS-FNO-EE-ADC(3) with {CIS(D) SS-FNOs} is 0.008 eV and is reduced to 0.007 eV with {the} perturbative correction. {The other statistical measures follow the same trend}. The error{-}distribution plot in Figure~\hyperref[fig:valence-theil]{4} shows that {truncating} the virtual space in the FNO basis leads to a slight {overestimation} of the excitation {energies} relative to the untruncated values. {SS-FNO-EE-ADC(3) with the CIS(D) density} shows a small positive deviation (ME $=$ 0.002 eV), consistent with the known tendency of CIS(D) to overshoot excitation energies.\cite{Loos_2025_QUEST}
\begin{figure}
    \centering
    \includegraphics[width=0.5\textwidth]{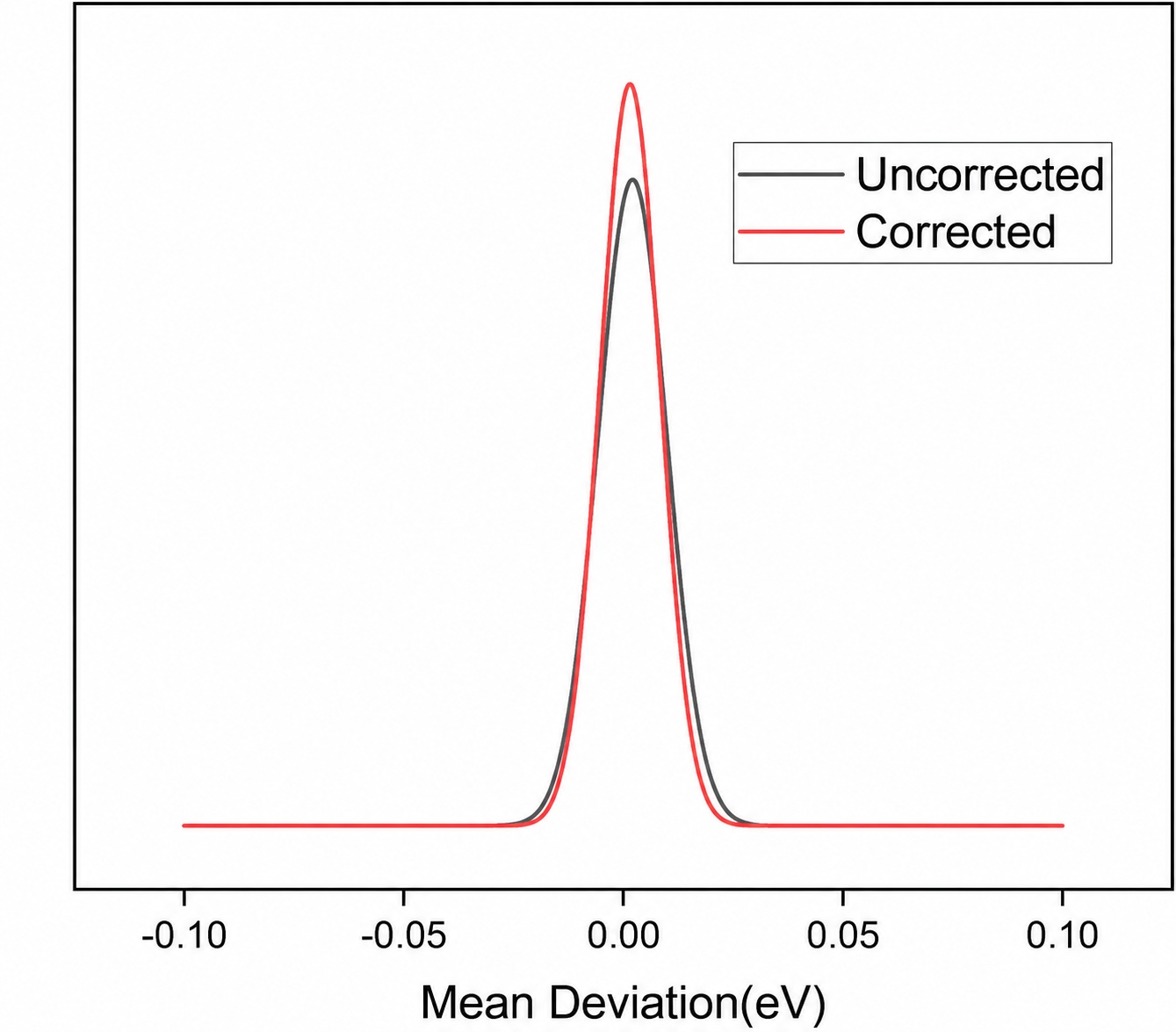}
    \caption{\label{fig:valence-theil} Error distributions {of the} uncorrected and corrected SS-FNO-EE-ADC(3) results for the valence excited states in the {Schreiber--Thiel} test set using CIS(D) densities. Errors are {measured relative to} canonical EE-ADC(3) {in} the TZVP basis set ($10^{-4}$ excited{-state} FNO threshold).
    }
\end{figure}

\begin{table}[h!]
\centering
\caption{Statistical {analysis} of SS-FNO-EE-ADC(3) excitation energies (eV) relative to canonical EE-ADC(3) {values} for the valence excited {states} {of the Schreiber--Thiel} test set ($10^{-4}$ excited{-state} FNO threshold).
}
\label{tab:thiel-tzvp}
\begin{tabular}{lcc}
\hline
{\textbf{Measure}} &
\textbf{Uncorrected} & \textbf{Corrected} \\
\hline
ME & 0.002 & 0.002 \\
{STDEV} & 0.008 & 0.007 \\
{MAXE} & 0.029 & 0.026 \\
RMSD & 0.008 & 0.007 \\
MAE & 0.006 & 0.005 \\
\hline
\end{tabular}
\end{table}

\subsubsection{Rydberg Test Set}
{To} evaluate the performance of the SS-FNO-EE-ADC(3) method for singlet Rydberg excited states, {we use} the Waterloo test set of Nooijen {and co-workers}.\cite{dutta2018exploring} The calculations {use Ahlrichs'} TZVP basis set, supplemented with {$5s4p4d$} diffuse functions {on a ghost atom at the center of mass}. {Untruncated} canonical EE-ADC(3) results serve as the reference{, and their} statistical analysis is presented in Table~\hyperref[tab:Rydberg-tzvp]{II}. {Figure~\hyperref[fig:rydberg]{5} shows that the truncated SS-FNO basis slightly underestimates the excitation energies}. The {maximum absolute error} is 0.044 eV with the CIS(D) density and decreases slightly to 0.043 eV with the perturbative correction. The RMSD and MAE {are} 0.012 and 0.009 eV{, respectively; the RMSD is unchanged by the perturbative correction, whereas the MAE decreases to} 0.008 eV. {Overall, the perturbative correction slightly improves both the magnitude and the spread of the error}. The mean error {with the perturbative correction} is $-0.003$ eV, {a slight underestimation that mirrors the behavior of the parent CIS(D) method, which tends to underestimate Rydberg excitation energies}.\cite{Loos_2025_QUEST}

\begin{figure}[h]
    \centering
    \includegraphics[width=0.5\textwidth]{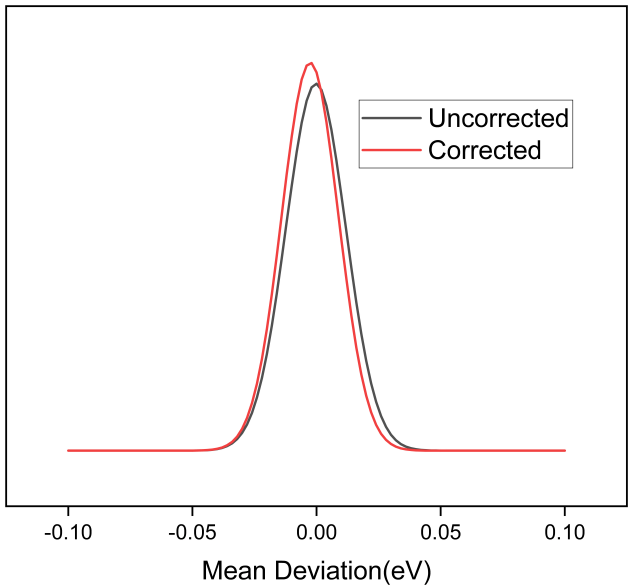}
    \caption{\label{fig:rydberg} Error distributions {of the} uncorrected and corrected SS-FNO-EE-ADC(3) results for the {Waterloo Rydberg} test set using CIS(D) densities. Errors are {measured relative to} canonical EE-ADC(3) {in} the TZVP basis set {(with additional $5s4p4d$ functions at the center of mass)} ($10^{-4}$ excited{-state} FNO threshold).
}
\end{figure}

\begin{table}[h!]
\centering
\caption{Statistical analysis of SS-FNO-EE-ADC(3) excitation energies (eV) relative to canonical EE-ADC(3) {values} for the Waterloo Rydberg test set ($10^{-4}$ excited{-state} FNO threshold).}
\label{tab:Rydberg-tzvp}
\begin{tabular}{lcc}
\hline
{\textbf{Measure}} &
\textbf{Uncorrected} & \textbf{Corrected} \\
\hline
ME & -0.000 & -0.003 \\
{STDEV} & 0.012 & 0.011 \\
{MAXE} & 0.044 & 0.043 \\
RMSD & 0.012 & 0.012 \\
MAE & 0.009 & 0.008 \\
\hline
\end{tabular}
\end{table}

\subsubsection{Charge-Transfer Test Set}
{TD-DFT and many approximate excited-state methods are known to struggle with charge-transfer (CT) states, so it is important to assess the SS-FNO-EE-ADC(3) method for this class of excitations. We use the benchmark set of Kozma and co-workers},\cite{kozma2020new} {which consists of ten dimers with low-lying CT states; the excitation energies are computed in the} cc-pVDZ basis set. {Untruncated} canonical EE-ADC(3) excitation energies are used as the reference{, and their} statistical analysis is presented in Table~\hyperref[tab:CT-cc-pVDZ]{III}. The {mean error} is $-0.026$ eV {with the CIS(D) density and improves to} $-0.011$ eV with the {perturbative} correction, while the maximum absolute error drops sharply from 0.095 eV to 0.050 eV. The other statistical measures improve similarly, so the perturbative correction markedly reduces both the magnitude and the spread of the error for CT states. The error{-}distribution plot in Figure~\hyperref[fig:charge-transfer]{6} {shows that the CT} excitation energies are underestimated {with the SS-FNO CIS(D) density, mirroring the behavior of CIS(D) itself for CT states}.
\begin{figure}
    \centering
    \includegraphics[width=0.5\textwidth]{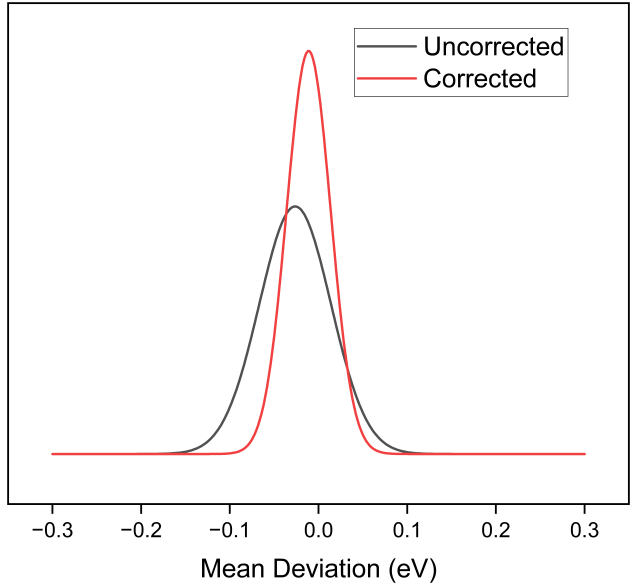}
    \caption{\label{fig:charge-transfer} Error distributions {of the} uncorrected and corrected SS-FNO-EE-ADC(3) results for the {charge-transfer} test set using CIS(D) densities. Errors are {measured relative to} canonical EE-ADC(3) {in} the cc-pVDZ basis set ($10^{-4}$ excited{-state} FNO threshold).
    }
\end{figure}
\begin{table}[h!]
\centering
\caption{Statistical {analysis} of SS-FNO-EE-ADC(3) excitation energies (eV) relative to canonical EE-ADC(3) {values} for the charge-transfer test set ($10^{-4}$ excited{-state} FNO threshold{)}.}
\label{tab:CT-cc-pVDZ}
\begin{tabular}{lcc}
\hline
{\textbf{Measure}} &
\textbf{Uncorrected} & \textbf{Corrected} \\
\hline
ME & -0.026 & -0.011 \\
{STDEV} & 0.040 & 0.025 \\
{MAXE} & 0.095 & 0.050 \\
RMSD & 0.047 & 0.026 \\
MAE & 0.029 & 0.018 \\
\hline
\end{tabular}
\end{table}

\subsection{Benchmarking on an Extended Test Set}
The method is further tested on the {QUEST}\cite{Loos_2025_QUEST} data set. Table~\hyperref[tab:quest_stats]{IV} presents the statistical error analysis of {the} excitation energies for the QUEST benchmark set computed {in} the aug-cc-pVTZ {basis set} at a FNO truncation threshold of $10^{-4}$. The mean error (ME) {is} very close to zero {in} all cases, indicating negligible systematic error relative to the canonical reference. The uncorrected SS-FNO calculation shows a ME of $-0.003$ eV, which shifts slightly to $-0.008$ eV {with} the correction{; the} SA-FNO approach yields ME {values} of 0.002 eV (uncorrected) and $-0.005$ eV (corrected). These {small} deviations indicate that both truncation schemes {preserve} the absolute excitation-energy scale with high accuracy.
\begin{table}
\caption{Statistical error analysis of SS-FNO- and SA-FNO-EE-ADC(3) excitation energies (eV) relative to canonical EE-ADC(3) values for the QUEST data set {in} the aug-cc-pVTZ basis set at a FNO threshold of $10^{-4}$.}
\label{tab:quest_stats}
\renewcommand{\arraystretch}{1.2}
\begin{tabular}{lcccc}
\hline\hline
 
& \textbf{Uncorrected} 
& \textbf{Corrected} 
& \textbf{Uncorrected} 
& \textbf{Corrected} \\
& \textbf{(SS)} 
& \textbf{(SS)} 
& \textbf{(SA)} 
& \textbf{(SA)} \\
\hline
ME    & -0.003 & -0.008 & 0.002 & -0.005 \\
STDEV & 0.012  & 0.009  & 0.016 & 0.007 \\
MAXE  & 0.026  & 0.031  & 0.036 & 0.032 \\
RMSD  & 0.013  & 0.013  & 0.016 & 0.009 \\
MAE   & 0.010 & 0.011 & 0.014 & 0.007 \\
\hline\hline
\end{tabular}
\end{table}
The SS-FNO calculations {show} a standard deviation of {about} 0.012 eV and {an} RMSD {of about} 0.013 eV, whereas the SA-FNO approach {shows a standard deviation of 0.016 eV and an RMSD of 0.016 eV (uncorrected), improving to 0.007 and 0.009 eV with the correction}. {The} mean absolute error (MAE) {for SA-FNO} decreases from 0.014 to 0.007 eV {upon adding the} perturbative correction, {whereas for SS-FNO the MAE is essentially unchanged (0.010 vs.\ 0.011 eV)}. The maximum {absolute} error (MAXE) remains {small} across all cases {(0.026--0.036 eV)}, and the perturbative correction does not significantly increase it. Overall, the state-specific and state-averaged formulations give similar excitation energies. However, the size of the virtual space {grows with the number of roots in the state-averaged scheme, which is not the case for the state-specific FNO approach}.\\
Figure~\hyperref[fig:common]{7} {shows scatter plots of the retained virtual space and the truncation error for the QUEST data set (aug-cc-pVTZ) in the SS-FNO-EE-ADC(3) method}.
The horizontal axis {is} the total number of virtual orbitals in the canonical reference calculation, the left vertical axis (blue circles) {is} the percentage of virtual orbitals retained after truncation, and the right vertical axis (red crosses) {is} the excitation-energy error introduced by the truncation. The {fraction} of retained virtual orbitals generally decreases with increasing system size, {with} a noticeable spread across {the QUEST molecules}. Smaller systems tend to retain a larger fraction of the virtual space, often exceeding 60--70\%, whereas larger systems typically retain {around} 40--55\%. Despite this {large} reduction, the truncation errors remain {small} throughout the data set, {most} below {about} 0.015 eV{, with} only a few outliers approaching 0.03 eV. {Thus} significant compression of the virtual space {is} achieved without {appreciable} loss in excitation-energy accuracy.\\

\begin{figure*}
    \centering
        \includegraphics[width=\linewidth]{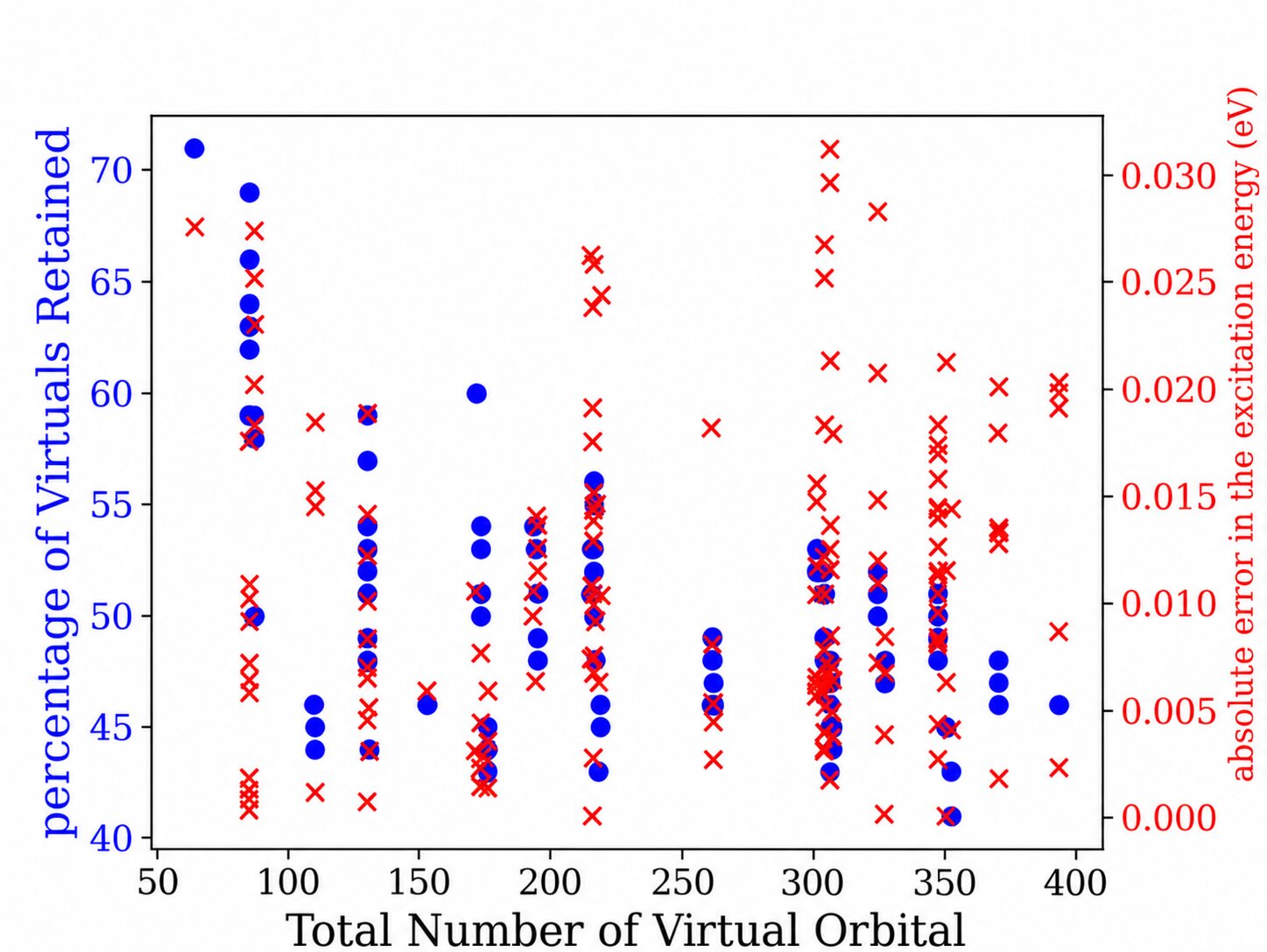}
    \caption{Percentage of virtual natural orbitals (VNOs) retained {(blue circles, left axis)} and absolute error in the excitation {energy} (eV) {relative to the canonical result (red crosses, right axis)}, as a function of the {number of canonical virtual orbitals}, for the {QUEST} test set {in} the aug-cc-pVTZ basis set {with the} SS-FNO-EE-ADC(3) method at {a} $10^{-4}$ {FNO} threshold.
    }
    \label{fig:common}
\end{figure*}
\subsection{Computational Efficiency and Application {to} a Large Molecule}
To assess the efficiency of our implementation of the SS-FNO-EE-ADC(3) method, we {computed the lowest excited state of a series of water clusters, $(\mathrm{H_2O})_n$ ($n=1$--5), in the aug-cc-pVTZ basis set with both the SS-FNO and the canonical EE-ADC(3) methods}. Figure~\hyperref[fig:timecalculation]{8} {shows the wall time versus} cluster size. The calculations were performed on a {dedicated workstation with two Intel Xeon Gold 5315 CPUs (16 cores total)} and 2 TB of RAM. {The wall time of the SS-FNO-EE-ADC(3) method grows more slowly with cluster size than that of the canonical method}.
\begin{figure}
    \centering
    \includegraphics[width=0.5\textwidth]{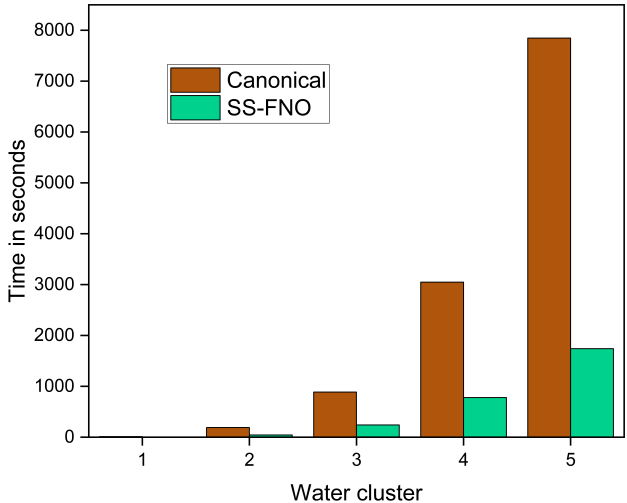}
    \caption{\label{fig:timecalculation} {Wall time of the} canonical EE-ADC(3) and SS-FNO-EE-ADC(3) {methods for water clusters $(\mathrm{H_2O})_n$ ($n=1$--5)} ($10^{-4}$ excited{-state} FNO threshold).}
\end{figure}
\\
To further {probe} the performance of the SS-FNO-EE-ADC(3) method, we {computed the first excitation energy of a medium-sized molecule, $N$-methyl-2,3-benzcarbazole (Figure~\hyperref[fig:big-molecule]{9}), which consists of 31 atoms and 122 electrons; its geometry was taken from Ref.~\cite{manna2025reducedcostequationmotion}}.
\begin{figure}
    \centering
    \includegraphics[width=0.5\textwidth]{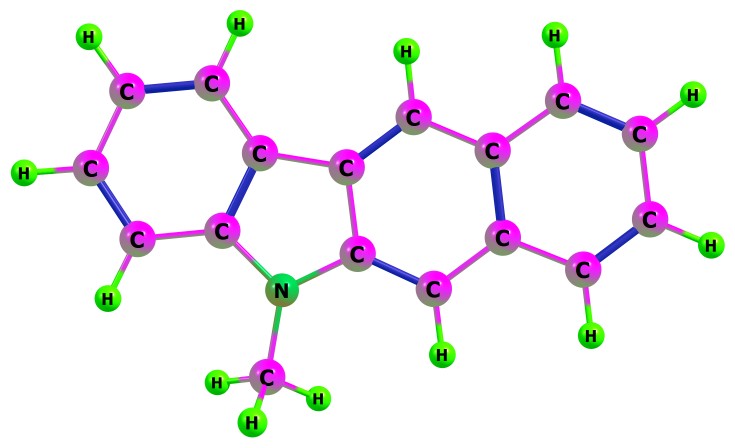}
    \caption{\label{fig:big-molecule} Structure of {$N$}-methyl-2,3-benzcarbazole.}
\end{figure}
{The aug-cc-pVTZ basis set gives a virtual space of 1066 orbitals. An FNO truncation threshold of $10^{-4}$ was used for both the ground and the excited states, leaving a truncated virtual space of 506 for the excited state. The resulting excitation energy is 3.68~eV, in good agreement with the previously reported EOM-CCSD value of 3.63~eV}.\cite{manna2025reducedcostequationmotion} {The entire calculation took 7~h 45~min, of which 4~min were spent on the Hartree--Fock step and 3~h on forming the EE-ADC(3) intermediates; each Davidson iteration took 54~min. For comparison, an SS-FNO-EOM-CCSD calculation on the same system took 1~day 12~h}.\cite{manna2025reducedcostequationmotion}

\section{Conclusion}
In this work, we have developed and implemented a low-scaling formulation of the algebraic diagrammatic construction method for excited states by combining tensor hypercontraction (THC), density fitting, natural auxiliary functions, and frozen natural orbital techniques. The primary objective was to overcome the steep computational scaling and memory bottlenecks {of} higher-order ADC methods{, particularly ADC(3), while} preserving their well-established accuracy for a wide range of excited-state problems. The SS-FNO approximation {generated from the CIS(D) one-particle reduced density gives systematic convergence of the excitation energy with the size of the virtual space and is controlled by a single threshold}. {Regenerating the integrals} in the state-specific natural orbital basis {is done efficiently with the density fitting approximation, which also reduces the storage requirement substantially}. The THC approximation significantly {reduces the cost of the four-external ladder term, whose formal scaling cannot be lowered by density fitting alone, and it combines well with the natural-orbital-based reduced-cost implementation}.
Extensive benchmarking shows that the proposed SS-FNO-EE-ADC(3) method delivers excitation energies with accuracy comparable to canonical ADC(3) across valence, Rydberg, and charge-transfer excited states. {We have also tested state-averaged natural orbitals; the SA-FNO-EE-ADC(3) method gives accuracy similar to the state-specific approach}. However, the {compression} offered by the state-averaged approach {becomes negligible when many roots are computed}. From a computational perspective, the combined use of THC, NAF, and SS-FNO leads to a substantial reduction in cost and a significantly slower growth of wall time with system size compared to canonical ADC(3). This enables excited-state ADC(3) calculations on molecular systems well beyond the sizes previously accessible, as demonstrated by applications to extended water clusters and a medium-sized organic chromophore with over one thousand virtual orbitals. The present work demonstrates that state-specific natural orbitals generated from the CIS(D) density are an effective way to reduce the cost of the ADC(3) method for excited states dominated by single excitations. Further cost reductions, including a formal reduction in scaling, can be obtained by combining natural-orbital techniques with local correlation approaches; work in this direction is in progress.

\section{Acknowledgments}
AKD, AM, and SC acknowledge {support} from IIT Bombay, {the} IIT Bombay Seed Grant (Project No. R.D./0517-IRCCSH0-040), {the} CRG (Project No. CRG/2022/005672) and MATRICS (Project No. MTR/2021/000420) {projects} of DST-SERB, CSIR-India (Project No. 01(3035)/21/EMR-II), {the} DST-Inspire Faculty Fellowship (Project No. DST/INSPIRE/04/2017/001730), and ISRO{, as well as the} IIT Bombay {supercomputing} facility and C-DAC {supercomputing resources (Param Smriti and Param Bramha) for computational time}. AM {acknowledges the} CSIR SRF Fellowship{;} SC {acknowledges the} Prime Minister's Research Fellowship (PMRF).

\bibliography{ref}

\end{document}